\documentclass[12pt,a4paper,twoside]{amsart}
\usepackage{amsmath,bbm,amssymb,amsxtra}
\usepackage{enumerate}
\allowdisplaybreaks
\usepackage{epsfig}
\usepackage{ae}
\theoremstyle{definition}

\newtheorem{lemma}{Lemma}[section]
\newtheorem{proposition}[lemma]{Proposition}

\newtheorem{theorem}[lemma]{Theorem}

\newtheorem{definition}[lemma]{Definition}

\newcommand{\prop}[1]{\begin{proposition}\label{#1}
\sl }
\newcommand{\eprop}{\end{proposition}}
\newcommand{\thm}[1]{\begin{theorem}\label{#1}
\sl }
\newcommand{\ethm}{\end{theorem}}
\newcommand{\lem}[1]{\begin{lemma}\label{#1}
\sl }
\newcommand{\elem}{\end{lemma}}

\newcommand{\defin}[1]{\begin{definition}\label{#1}
\sl }
\newcommand{\edefin}{\end{definition}}

\newcommand{\beqno}{\begin{eqnarray*}}
\newcommand{\eeqno}{\end{eqnarray*}}
\newcommand{\beqla}[1] {\begin {eqnarray}\label{#1}}
\def\eeq {\end {eqnarray}}
\newcommand{\beq}{\begin {eqnarray}}

\newcommand{\E}{\mathbb{E}}

\newcommand{\R}{\mathbb{R}}

              \def\CL{{\mathcal L}}
              \def\CO{{\mathcal O}}

\def\qq{ \begin{eqnarray} }
\def\qqq{ \end{eqnarray} }
\def\rr{ \begin{equation} }
\def\rrr{ \end{equation} }

\newcommand{\NJ}{{{\bf J}}}

\def\qq{ \begin{eqnarray} }
\def\qqq{ \end{eqnarray} }

\newcommand{\no}{\noindent}

\newcommand{\hf}{{_1\over^2}}

\newcommand{\BbbN}{\mathbb{N}}
\newcommand{\BbbZ}{\mathbb{Z}}

\newcommand{\BbbT}{\mathbb{T}}

\begin{document}
\title[Origins of Diffusion]
{Origins of Diffusion}

\author[A. Kupiainen]{Antti Kupiainen$^1$}
\address{University of Helsinki, Department of Mathematics and Statistics,
         P.O. Box 68 , FIN-00014 University of Helsinki, Finland}
\email{antti.kupiainen@helsinki.fi}
\thanks{Supported by European Research Council and Academy of Finland}%


\begin{abstract}
We consider a dynamical system consisting of subsystems indexed by a lattice. Each subsystem
has one conserved degree of freedom ("energy") the rest being uniformly hyperbolic. The subsystems are weakly coupled together so that the sum of the subsystem energies remains conserved.
We prove that the long time dynamics of the subsystem energies is diffusive.
\end{abstract}

\maketitle


\label{se:intro} 

\section{Diffusion from conservative dynamics}
\label{se:intro} 
One of the fundamental problems in deterministic dynamics is to understand the
microscopic origin of dissipation and diffusion. On a microscopic level a physical system such as
a fluid or a crystal can be described by a Schr\"odinger or a Hamiltonian dynamical system with a macroscopic
number of degrees of freedom. Although the microscopic dynamics is reversible in time one
expects 
dissipation to emerge in large spatial and temporal scales e.g. in the form
of diffusion of heat or concentration of particles. 

To fix ideas, consider a Hamiltonian dynamical system i.e. a Hamiltonian flow on a
symplectic manifold $M$. For the present purpose it suffices to consider $M=\R^{2n}$
with position and momentum coordianates $q,p\in\R^n$. The Hamiltonian flow $\phi_t\in{\rm Diff}M$
generated by the vector field $(\partial_pH,-\partial_qH)$ where $H:M\to\R$ is
the Hamiltonian or energy function preserves the energy
$$H\circ\phi_t=H$$
i.e. the flow preserves the constant energy sets $M_E=\{(q,p):H(q,p)=E\}$.

On the other hand, the simplest diffusion process is given by the heat equation
\qq
\partial_tE(t,x)=\kappa\Delta E(t,x)
\label{heat}
\qqq
and the associated semigroup $\psi_t=e^{\kappa t\Delta }$. Unlike for  the reversible $\phi_t$
where $\phi_{-t}=\phi_t^{-1}$, $\psi_t$ has no inverse and describes dissipation.
Physically, the energy function $E(t,x)$ describes a macroscopic energy density
i.e. a coarse grained function of  microscopic dynamical variables, the positions and
momenta of the underlying Hamiltonian dynamics. The question we wish to
pose is how does this dissipative dynamics $\psi_t$ arise from the conservative one $\phi_t$. 

A concrete physical system where diffusion occurs is a fluid. In classical mechanics
this is microscopically modeled by a Hamiltonian system whose flow gives the trajectories of
the fluid particles $(q_i(t),p_i(t))\in\R^3\times\R^3$, $i=1\dots N$. A typical Hamiltonian function is given by
\qq
H(q,p)=\sum_i{p_i^2\over 2m}+\sum_{i j}V(q_i-q_j)
\label{fluid}
\qqq
consisting of the kinetic energy of the particles of mass $m$ and a pair potential energy of interaction
of the particles. Let the energy of the i:th particle be defined as 
\qq
e_i={p_i^2\over 2m}+\hf\sum_{j\neq i}V(q_i-q_j)
\label{ei}
\qqq
so that $H=\sum_i e_i$. We can then define the energy density as the distribution
\qq
E(t,x)=\sum_i e_i\delta_{q_i(t)}(x)
\label{E}
\qqq
where $\delta_{q}$ is the Dirac mass at $q$. Since $\int E(t,x)dx=\sum_ie_i=H$ and $\dot H=0$
one concludes
\qq
\dot E(t,x)=\nabla\cdot J(t,x)
\label{conservation}
\qqq
for a certain distribution, the energy current, depending on $q(t),p(t)$.  Eq. (\ref{conservation}) is
a local conservation law deduced from the global energy conservation. In the case of the fluid,
there are two other similar local conservation laws related to global momentum and particle number 
conservation laws. This leads to a richer set of macroscopic laws in the case of the fluid than the
diffusion equation for the energy (in particular these include the Navier-Stokes equations).

\section{Coupled oscillators}

Thus, to understand the origins of diffusion one should look for systems with just one local conservation
law eq. (\ref{conservation}). 
There has been a lot of  work in recent years around these questions
in the context of {\bf coupled dynamics} i.e. dynamical systems
consisting of elementary systems indexed
 by 
a $d$-dimensional lattice $\mathbb{Z}^d$. The total
energy $E$ of the system is a sum $\sum_{x}E_x$
of energies $E_x$ which involve the dynamical
variables of the system at lattice site $x$ and
nearby sites.
The physical situation to keep in mind is then thermal conduction in
a crystal lattice (i.e. a solid). 

Two types of systems have been considered. In the first case at each lattice
site we have an oscillator and the oscillators at neighboring sites are
coupled together. Typically one considers the system where the forces
are weakly anharmonic. In the second case at each lattice site one
puts a chaotic system and weakly couples the neighboring systems.
Let us start with the former case.

The setup resembles that of the fluid above, but now the  "particle" positions $q_x$ are indexed by the lattice, 
$x\in\Lambda\subset \BbbZ^d$ where $\Lambda$ is a finite subset, say a cube,
and they describe the deviation of an atom from its equilibrium position at $x$. A simple classical
mechanical model for this is a system of coupled oscillators
\qq
H_\Lambda(q,p)=\sum_{x\in\Lambda}({p_x^2\over 2m}+U(q_x))+\sum_{|x-y|=1}V(q_x-q_y)
\label{crystal}
\qqq
where $U$ is a {\it pinning} potential which we assume tending to infinity as $|q|\to\infty$. The potential $V$
describes interaction of the atoms in nearest neighbor lattice sites and is taken attractive. A challenging 
model is obtained already by  taking 
\qq
V(q)=q^2,\ \ \ U(q)=q^2+\lambda |q|^4
\label{vu}
\qqq and further simplifying by taking $q_x\in \R$
instead of $\R^d$. Then a lattice version of eq. (\ref{conservation}) holds with the current given by
\qq
J_\mu(x)=-\hf  (p_{x+\mu}+p_{x})V'(q_{x+\mu}-q_{x}).
\label{current}
\qqq

In what sense should we expect the conservative dynamics (\ref{conservation}) 
give rise to a diffusive one as in eq. (\ref{heat})? The answer is that this should
happen for {\it typical} initial conditions $(q(0),p(0))\in M_\Lambda$ with respect to a specific measure on the phase space  $M_\Lambda:= \R^{2|\Lambda|}$ and under a proper
{\it scaling limit}. 

Recall first that the Hamiltonian dynamics preserves the Lebesgue measure $m_\Lambda$
on  $M_\Lambda$. Since also $H_\Lambda$ is preserved so is the Gibbs measure
(or {\it equilibrium measure})
$$
\mu_{\beta\Lambda}=Z_\Lambda^{-1}e^{-\beta H_\Lambda}m_\Lambda
$$
where $\beta>0$ as well as its (thermodynamic) limit $
\mu_{\beta}=\lim_{\Lambda\to\BbbZ^d}\mu_{\beta\Lambda}$.  Let us now replace the (inverse) temperature parameter $\beta$
by a spatially varying one. Let $b\in  C_0^\infty(\R^d)$ and $\beta>\|b\|_\infty$.
Write as in the fluid case 
$$H_{\Lambda}=\sum_{x\in\Lambda}e_x$$, $e_x$
describing the energy contributed by the oscillator at $x$. Pick a
scaling parameter $L\in\BbbN$ and set $\beta_L(x)=\beta+b(x/L)$. Let
$\mu^{(L)}$ be the thermodynamic limit of the measure
 $$
Z_{L,\Lambda}^{-1}e^{-\sum_{x\in \Lambda}\beta_L(x)e_x}m_\Lambda.
$$
Construction of this limit poses no problems if $\lambda\geq 0$ in eq. (\ref{vu})
is small enough. $\mu^{(L)}$ is  {\it not} invariant under the dynamics
which maps it to $\mu^{(L)}_t=\mu^{(L)}\circ \phi_t^{-1}$. However,
one expects that as $t\to\infty$ there is {\it return to equilibrium} i.e.
$\mu^{(L)}_t\to \mu_{\beta}$. The diffusion equation is expected
to govern this process in the following sense.

Let $f\in C_0^\infty(\R_+\times\R^d)$ and consider the
random variables
\qq
e_{L}(f)=L^{-d-2}\sum_{(t,x)\in\BbbZ_+\times\BbbZ^d}f(t/L^{-2},x/L)e_x(q(t),p(t)).
\label{eL}
\qqq
The statement of the {\it hydrodynamic limit} is then: with probability
one in the sequence of measures $
\mu^{(L)}$,  $
e_{L}(f)$ converges to $\int f(t,x)E(t,x)dtdx$ where $E$ is the
solution to the nonlinear diffusion equation
\qq
\partial_t E=\nabla\cdot (\kappa(E)\nabla E)
\label{hydro}
\qqq
where $\kappa(E)$ is a smooth positive function. The initial condition $E(0,\cdot)$ is determined by the
function $b$. Thus upon coarse graining and scaling the equation 
(\ref{conservation}) turns to eq. (\ref{hydro}), {\it almost surely}
in the initial conditions of the underlying microscopic variables.

The proof of the hydrodynamic limit in our model is beyond present mathematical
techniques. The existing techniques require
the presence of plenty of noise in the system. A simpler problem would be
to establish the {\it kinetic limit}. This is a {\it weak anharmonicity} limit.
We replace $\lambda$ in eq.  (\ref{vu}) by $\lambda/\sqrt{L}$ and 
and consider the measures $\mu^{(L)}_{Lt}$. As $L\to\infty$ we expect
these measures to become gaussian whose covariance upon spatial
scaling satisfies a Boltzman equation. More precisely, denote $(q_x,p_x)$
by $\phi(x)$. Then it is conjectured that
\qq
\lim_{L\to\infty}\int \phi(Lx+y)\phi(Lx-y)\mu^{(L)}_{Lt}(d\phi)=G(t,x,y)
\label{kin}
\qqq
exists and the Fourier transform of $G(t,x,y)$ in $y$, $\hat G(t,x,k)$
satisfies the so called phonon Boltzman equation
\qq
\partial_t\hat G(t,x,k)+\nabla\omega(k)\cdot\nabla \hat G(t,x,k)=I(\hat G(t,x,\cdot))
\label{bol}
\qqq
where $I$ is a nonlinear integral operator and $\omega(k)^2$ is
the Fourier transform of the lattice operator
$2(-\Delta+1)$, see \cite{spohn}. Proof
of these statements is still open and a considerable challenge (for
some progress see \cite{LS}). Derivation of a hydrodynamic
equation of the type (\ref{hydro}) from the Boltzman equation
 (\ref{bol}) has been carried out \cite{BKbol}, see also  \cite{BKclos} where
 an attempt to go beyond the kinetic limit was carried out.

\section{Coupled chaotic flows}

A second class of models deals with a complementary situation of {\it weakly coupled
chaotic} systems  \cite{buni}, \cite{meja}, \cite{ey}. The setup is as follows.
Let $(M,H)$ be a
Hamiltonian system i.e. $M$ is a symplectic manifold and $H:M\to\R$.
Let, for each $x\in\BbbZ^d$ $(M_x,H_x)$ be a copy of $(M,H)$.  Let
$h:M\times M\to \R$ and for each $x,y\in\BbbZ^d$, $|x-y|=1$ let
$h_{xy}:M_x\times M_y\to \R$ be a copy of $h$. Let
$\Lambda\subset\BbbZ^d$ be finite and $M_\Lambda=\times_{x\in\Lambda}M_x$.
The coupled flow is the one on $M_\Lambda$ generated by the Hamiltonian
\qq
H_\Lambda=\sum_{x\in\Lambda}H_x+\sum_{|x-y|=1}\lambda h_{xy}.
\label{cham}
\qqq
Of course, the coupled oscillators of the previous section are of this
form. There, the system $(M,H)$ is integrable, and the diffusive dynamics 
is the consequence of coupling and anharmonicity. In the present discussion
we wish to take $(M,H)$ chaotic. Examples are Anosov systems
or billiard systems. E.g. in the former case the flow on $M$ generated by
$H$ has ${\rm dim}M-2$ non-zero Lyapunov exponents and two
vanishing ones corresponding to 
the Hamiltonian
vector field and  $\nabla H$. 

When the coupling parameter $\lambda$ is zero $(M_\Lambda,H_\Lambda)$
has $2|\Lambda|$ vanishing Lyapunov exponents. For $\lambda\neq 0$
one expects that for a large class of perturbations $h$ the only
constant of motion is $H_\Lambda$ and the system has only two
vanishing exponents. However, zero should be near degenerate
for the Lyapunov spectrum and these  long time scale motions should
be at the origin
to diffusion in the $\Lambda\to\BbbZ^d$ limit. 

Rigorous results on such Hamiltonian systems are rare: in  \cite{buni}
ergodicity is proved in a one dimensional model. However,
it seems very difficult to get hold of the Lyapunov spectrum
and it is far from obvious how such knowledge would turn into
a proof of diffusion in these systems.  I want to argue that a
more fruitful approach is to study the local energy conservation law
(\ref{conservation}) and try to show that the chaotic degerees
of freedom act there like a noise that redistributes locally the
energy. To probe such an idea it is useful to turn to
a discrete time version of our model i.e. to study iteration
of a map rather than a flow.

\section{Coupled chaotic maps}

A discrete time version of the coupled flow setup of the
previous section is called a
Coupled Map Lattice (CML).  Now the local dynamical system is a pair $(M,f)$
where $M$ is a manifold and $f:M\to M$. Again for each $x\in\BbbZ^d$ $(M_x,f_x)$ is a copy of $(M,f)$ and $(M_\Lambda, f_\Lambda)$ with $f_\Lambda=\times_{x\in\Lambda}f_x$
is the product dynamics. The CML
 dynamics is a suitable local perturbation  of the product dynamics.
 
 Our choice of $M$ and $f$ is motivated by the  coupled chaotic flows
discussed before.
 A discrete time version (say given by a Poincare map) of a billiard or Anosov
 flow  has one 
vanishing  Lyapunov exponent  corresponding to the conserved energy
 and the remaining ones nonzero.
We model such a situation by taking for the local dynamics the manifold
of form
 $M= \mathbb{R}_+\times N$ with $N$ another manifold. Let us denote
 the variables at the lattice site $x\in\BbbZ^d$ by $(E(x),\theta(x))\in \mathbb{R}_+\times N$.
We call the non-negative variables $E$ energy and postulate them
to be
conserved under  the local  dynamics:
\qq
(E(x),\theta(x))\to (E(x),g(\theta(x), E(x)))
\label{uc}
\qqq
for each $x\in\mathbb{Z}^d$.

$\theta\in N$ are the fast, chaotic variables. In the billiard case 
the dynamical system $\theta\to g(\theta, E)$ is uniformly hyperbolic for any fixed E.
We will model this situation by taking $g(\theta, E)=g(\theta)$
a fixed chaotic map, independent of $E$. Examples are 
$N=\BbbT^1=\mathbb{R}/\mathbb{ Z}$ and $g$  an
expansive circle map, e.g $g(\theta)=2\theta$ and 
$N=\BbbT^2=\mathbb{R}^2/\mathbb{ Z}^2$ and $g$  a
hyperbolic toral automorphism.  

We should stress that the $E$ independence is the
most serious simplification in this setup.  In a realistic Hamiltonian system, such as the billiards the $E$ dependence
of $g$ can not be ignored. Indeed, it is obvious that as $E\to 0$ the Lyapunov exponents
of $g(\cdot,E)$ also tend to zero since $E$ sets the time scale.

The CML dynamics is a perturbation of the local dynamics (\ref{uc}).
Let us use the same notation $(E,\theta)\in M_\Lambda =\mathbb{R}_+^\Lambda
\times N^\Lambda$. Then $F:M_\Lambda\to M_\Lambda$ is written as
\qq
F(x,E,\theta)= (E(x)+f(x,E,\theta),g(\theta(x))+h(x,{\theta})).
\label{uc1}
\qqq
Here $f$ and $h$ are small local functions of $(E,\theta)$ i.e. they depend weakly on  
 $(E(y),\theta(y))$ for $|x-y|$ large as we will specify later.
 
  $f$ is however constrained by the requirement that the {\it total energy} $\sum_{x}E(x)$
 is conserved. This follows if
 $$
\sum_{x}f(x,E,\theta)=0
$$
 for all $E,\theta$. A natural way to guarantee this is to 
 consider a "vector field" $\NJ(x)=\{ J^\mu(x)\}_{\mu=1,\dots, d}$  and take
  \qq
f(x,E,\theta)=(\nabla\cdot \NJ)(x,E,\theta):=\sum_\mu (J^\mu(x+e_\mu, E,\theta)-J^\mu(x,E,\theta))
\label{current}
\qqq
With these definitions we arrive at the time evolution
\qq
E(t+1,x)&=&E(t,x)+\nabla\cdot \NJ(x,E(t),\theta(t))\label{Edyn}\\
\theta(t+1,x)&=&g(\theta(t,x))+h(x,\theta(t))).\label{thdyn}
\label{dyn}
\qqq
Note that (\ref{Edyn}) is a natural discrete space time  version of  (\ref{conservation}).
Let us discuss this iteration from a general perspective before making
more specific assumptions of the perturbations.

\section{Fast Dynamics}

The iteration (\ref{thdyn}) of the chaotic variables is autonomous. We shall
assume the perturbation $h$ is $C^1$ with the following locality
property
 \qq
|\partial _{\theta(y)}h(x,\theta)|\leq \epsilon e^{-a|x-y|}
\label{l1}
\qqq
and H\"older continuity property
 \qq
|\partial _{\theta(y)}h(x,\theta)-\partial _{\theta(y)}h(x,\theta')|\leq \epsilon\sum_z e^{-a(|x-y|
+|x-z|)}|\theta(z)-\theta'(z)|.
\label{l2}
\qqq
These properties guarantee \cite{cml} that the $\theta$-dynamics is {\it space-time
mixing}. This means that the dynamics is defined in the $\Lambda\to\BbbZ^d$
limit and it has a unique Sinai-Ruelle-Bowen measure $\mu$ 
on the cylinder sets of $N^{\BbbZ^d}$ which satisfies
 \qq
\E(F(\theta(t,x))G(\theta(0,x)))- \E(F(\theta(t,y)) \E G(\theta(0,y))
\leq C e^{-c(t+|x-y|)}
\label{stc}
\qqq
for H\"older continuous functions $F$ and $G$.  Here $\E$ denotes expectation in $\mu$.

We conclude that
sampling $\theta(0,\cdot)$ with $\mu$ makes
$\theta(t,x)$ {\it random variables} which are exponentially weakly correlated at distinct
space time points. Therefore $\theta_x(t)$ acts as a {\it random environment} for the slow variable dynamics (\ref{Edyn}).

\section{Quenched diffusion}

The previous discussion shows that we can view  the 
current $ \NJ(x,E,\theta(t))$ in the 
slow variable dynamics (\ref{Edyn}) as a random field $ \NJ(t,x,E)$
which is exponentially weakly correlated in space and time. 
  We may thus rephrase the problem of deriving diffusion in deterministic dynamics
 as that of {\it quenched diffusion in random dynamics}. We want to show
 that the random dynamical system
 \qq
E(t+1,x)&=&E(t,x)+\nabla\cdot \NJ(t,x,E(t)):=\Phi(t,x,E(t))\label{rdyn}
\qqq
has a diffusive hydrodynamical limit  {\it almost surely} with respect to the
SRB measure $\mu$. 
Let us inquire how this should come about and then list
the assumptions we need for the actual proof.

Consider first the 
{\it annealed} problem, i.e.  averaged equation (\ref{rdyn}): 
$$
E_x(t+1)-E_x(t)=\nabla\cdot\E [J(t,x,E(t))]
:=\nabla\cdot {\mathcal J}(x,E(t)).
$$
where, by stationarity of $\mu$, ${\mathcal J}$ is time independent.
Supposing that $h$ and $\NJ$ have natural symmetries under lattice
translations and rotations we infer that ${\mathcal J}$ vanishes at constant $E$
and then locality assumptions of the type we assumed for $h$
imply
$${\mathcal J}(x,E)=\sum_y\kappa (x,y,E)\nabla E(y).$$
Hence the annealed dynamics is a {\it discrete nonlinear diffusion}
$$
E(t+1)-E(t)=\nabla\cdot \kappa( E(t))\nabla E(t)
$$
provided the diffusion matrix $ \kappa( E(t))$ is positive. 

Let now
$$\beta(t,x,E(t))=J(t,x,E(t))- {\mathcal J}(x,E(t))$$
be the fluctuating part. Then slow dynamics becomes
$$
E(t+1)-E(t)=\nabla\cdot \kappa( E(t))\nabla E(t)
+\nabla \cdot \beta(t,E(t))
$$
$$\E\ \beta(t,E)=0$$
i.e. a {\it nonlinear diffusion} 
with {\it a random drift}. In a physical model one would expect  $\kappa( E(t))$ to be
{positive}  although not necessarily uniformly in $E$. If furthermore
 $\beta$ turned out to be a small perturbation quenched diffusion might be provable.
 In what follows we will make such assumptions and then indicate
 how to establish diffusion.
 
 Before stating the assumptions let us make one more
 reduction. It is reasonable to assume $E=0$ is preserved by the dynamics.
 This then implies  $\beta(t,0)=0$. Let us study the linearization at $E=0$:
  \qq
E(t+1)-E(t)=\nabla\cdot \kappa( 0)\nabla E(t)
+\nabla \cdot (D\beta(0,t)E(t))
\label{lin}
\qqq
or, in other words
 \qq
E_x(t+1)=\sum_{y}p_{xy}(t)E_{y}(t)
\label{lin}
\qqq
with 
$$\sum_xp_{xy}(t)=1.$$
Since $E\geq 0$ we have $p_{xy}\geq 0$ i.e. $p_{xy}(t)$ are 
{\it transition probabilities of a random walk}. $p_{xy}(t)$  is space and time dependent
and random i.e. it defines a
{\it  random walk in random environment}.

\section{Random walk in nonlinear random environment}

 Consider a random walk defined by the transition probability matrix $p_{xy}(t)$ 
 at time $t$. $p(t)=p(t,\omega)$ is a taken random defined on some probability
 space $\Omega$. We suppose the law of $p$ is invariant under translations in
 space and time. Define
  \qq
\|E\|:=\sup_x|E(x)|(1+|x|)^{d+a}
\label{norm}
\qqq
for some $a>0$.
 Let, at $t=0$, $\|E\|<\infty$.
 We say the walk  defined by $p$ is has a diffusive scaling limit if there
 exists $C,\kappa$ such that {\it almost surely} in $\omega$
  \qq
\lim_{L\to\infty}\|L^dE(L^2t,L\cdot)-Ct^{-d/2}E_\kappa^*(\cdot/\sqrt{t}\|= 0
\label{sl}
\qqq
where 
$E_\kappa^*(x)= e^{-x^2/4\kappa}$. In other words
$$L^dE(L^2t,Lx)\sim Ct^{-d/2}e^{- x^2/{ 4\kappa t}}$$
as
$ L\to\infty$.

We prove this for a {\it non-linear perturbation} of RWRE.  Let us state
the assumptions for the random dynamical system eq. (\ref{rdyn}).
We assume $\Phi$ is $C^2$ in $\|E\|_1<\delta$ and satisfies

\vskip 2mm

\no {\it Positivity:} $\Phi(E)\geq 0$ for $E\geq 0$.

\vskip 2mm

\no {\it Conservation law:} $$\sum_x\Phi(t,x,E)=\sum_xE_x$$

\vskip 2mm

\no {\it Weak nonlinearity:} $$|{\partial^2\Phi(t,x,E)\over\partial E_y\partial E_z}|\leq e^{-|x-y|-|x-z|}$$
Write the average map
$$\E \Phi(t,x,E)=\sum_y T(x-y)E_y+ o(E).$$ 

\vskip 1mm

\no {\it Ellipticity:}  $T$ generates a diffusive random walk on $\BbbZ^d$.

\vskip 2mm

\no Write
$$
\Phi(t,x,E)
-\E\Phi(t,x,E):=\nabla\cdot b(t,x,E).$$

\vskip 1mm

\no {\it Weak correlations.}   Assume
 \qq
b(t,x,E)=\sum_{A\subset {\BbbZ^d}\times [0,t]}b_A(t,x,E)
\label{bsum}
\qqq
with 
$$|b_A(t,x,E)|\leq \epsilon e^{-d((x,t)\cup A)}$$
and $b_A$, $b_B$ are {\it independent } if $A\cap B=\emptyset$.

\vskip 2mm

\no {\bf Remark.} A representation of the form (\ref{bsum}) arises
from the model we have discussed above with the proviso that
$b_A$, $b_B$ are {independent }  only in the case the $\theta$
dynamics is local, i.e. $h=0$ in eq. (\ref{thdyn}). For the
general $h$ there is weak dependence that can be handled.

\vskip 2mm

\begin{theorem}
 Under the above assumptions and $\delta$, $\epsilon$ small enough
  the random dynamical system
$\Phi_t$ is diffusive, almost surely in $\omega$.
\end{theorem}

\section{Renormalization group for random coupled maps}

The proof of Theorem 7.1. \cite{BK2} is based on a renormalization group 
method introduced in  \cite{BK3} and  \cite{BK4}. Let us introduce the
{scaling}  transformation $S_L$:
 \qq
(S_LE)(x)=L^dE(Lx).
\label{scaling}
\qqq
where $L>1$. Fix $L$ and define, for each $n\in\BbbN$, {\it renormalized energies}
$$
E_n(t)=S_{L^{n}}E(L^{2n}t).
$$
We can then rephrase the scaling limit (\ref{sl}) as
$$
\lim_{n\to\infty}L^{nd}E(L^{2n}t,{L^n x})=\lim_{n\to\infty}E_n(t,x).
$$
$E_n(t)$  inherits dynamics from $E$. We will call this
the {\it renormalized dynamics}:
$$
E_n(t+1)=\Phi_n(t, E_n(t)).
$$
Explicitely we have
 $$\Phi_n(t)=S_{L^{n}}(\Phi(L^{2n}t+L^{2n}-1)\circ\cdot\cdot\cdot\circ \Phi(L^{2n}t))S_{L^{-n}} .$$
The dynamics changes with scale as 
$$\Phi_{n+1}={\mathcal R}\Phi_n$$ 
with
$$
{\mathcal R}\Phi(t,\cdot)
=S_L\Phi(t_{L^2})\circ\cdot\cdot\cdot\circ
\Phi(t_1)S_{L}^{-1}
$$
with $t_1=L^2t$ and $t_{L^2}=L^2(t+1)-1$.
\vskip 2mm

$\mathcal R$ is the the {\it Renormalization group flow} in a space of random
dynamical systems.
We prove: {almost surely} 
the renormalized maps converge
$$
{\mathcal R}^nf\to f^*
$$
where the fixed point is {\it nonrandom and linear}:
$$
 f^*(E)= e^{\kappa\Delta}E.
$$
Moreover, the renormalized energies converge almost surely to the fixed point
$$\|E_n(t,\cdot) -{C\over t^{d/2}}E_\kappa^*(\cdot/\sqrt{t})\|\to 0$$
which is the diffusive scaling limit. 

These results may be summarized by saying that both the randomness
and the nonlinearity are {\it irrelevant} in the RG sense. Let us
 finish by sketching the reasons for this.
 
We start by considering the linear problem
$$
D\Phi(t,x,0)E=\sum_{y}p_{xy}(t)E_{y}.
$$
Then $DR\Phi=p'$ with
$$
p'(t)_{xy} = L^d(p({L^2(t+1)-1})\dots p(L^2t))_{LxLy}.
$$
Write 
$$p_{xy}(t)=T(x-y)+\nabla_y\cdot c_{xy}(t)$$
 with $\E p=T$ and $\E c=0$. Then, for
$p'=T'+\nabla c'$ we get
 \qq
T'(x-y)=L^dT^{L^2}(Lx-Ly)+r(x-y)
\label{tite}
\qqq
where $r$ is an expectation of a polynomial in $c$. For the
noise we get
 \qq
\nabla_x c'_{xy}
 = L^{d}\sum_{t=1}^{L^2}\sum_{uv}
T^t(Lx-u)
\nabla_u c_{uv}(t)T^{L^2-t-1}(v-Ly)+\gamma_{xy}.
\label{bite}
\qqq
where $\gamma$ involves quadratic and higher order polynomials in $c$. 

Ignoring first $r$ 
 we get for the
average flow $$T_{n}= L^{nd}T^{L^{2n}}(L^n\cdot)$$ i.e. $$\hat T_n(k)=\hat T^{L^{2n}}(k/L^n).$$
Write 
$\hat T(k)=1-ck^2+o(k^2)$. Then as $n\to\infty$:
$$\hat T_n(k)\to e^{-ck^2}$$
explaining the fixed point.

Similarly, ignoring $\gamma$ the noise is driven by the linear map
$$
 \CL c_{xy}(0)
 = L^{d-1}\sum_{t=1}^{L^2}\sum_{uv}
T^t(Lx-u)
c_{uv}(t)T^{L^2-t-1}(v-Ly).
$$
The variance of $ \CL c$ contracts:
 \qq
\E(\CL c)^2\sim L^{-d}Ec^2.
\label{var}
\qqq
The intuitive reason behind this is the following. Take e.g. $x=y=0$. For $t$ of order $L^2$, 
$T^t(Lx-u)\sim L^{-d}e^{-|x-u/L|}$. Hence the $u$ and the $v$ sums are
localized in an $L$ cube at origin.
Since $c_{uv}(t)$ has exponential decay in $|u-v|$ 
 \qq
 \CL c_{00}(0)\sim
  L^{-d-1}\sum_{t=1}^{L^2}\sum_{|u|<L}
c_{uu}(t).
\label{line}
\qqq
Since correlations
of $c$ decay exponentially in space and time (\ref{line}) is effectively a sum
of $L^{d+2}$ independent random variables of variance $L^{-2d-2}(\E c)^2$
thus leading to (\ref{var}). 

Taking into account the corrections $r$ and $\gamma$ in (\ref{tite}) and (\ref{bite})
we conclude that the  variance contracts as
$$\E(c_n)^2\sim \epsilon_n= L^{-nd}\epsilon.$$
The iteration of the mean becomes
 \qq
T_{n+1}=L^dT_n^{L^2}(L\cdot)+\CO(\epsilon_n).
\label{titen}
\qqq
The fixed point is the same but the $\CO(\epsilon_n)$ renormalizes the
diffusion constant $\kappa$ at each iteration step (less and less as $n\to
\infty$).

There is a problem however once we try to make  this perturbative
analysis rigorous. Deterministically the noise is {\it relevant}: from (\ref{line})
we see that $\|\CL c\|_\infty$
can be as big as 
$
\CO( L)\| c\|_\infty
$. This means that there are unlikely events in the environment where the random
walk develops a drift. 
We write
$$|c_{n}(t,,E)|\leq  L^{N_n(x)-bn}.$$
Then $N_n(x)$ can be (very) large, but with (very) small probability:
$${\rm Prob}(N_n(x)>N)\leq e^{-KN}$$
with $K$ large.

Finally, to control the {\it nonlinear} contributions to $\Phi_n$ we
show that  the second
derivative $D^2_E\Phi$ is irrelevant in all dimensions due to
the scaling of $E$:
$$
{\mathcal R}\Phi(t,x,E)=L^d(\Phi(t_{L^2})\circ\cdot\cdot\cdot\circ
\Phi(t_1))(Lx,L^{-d}E(\cdot/L)).
$$

\section{Towards Hamiltonian systems}

The coupled map lattices we have discussed are an alternative
microscopic model with a local conservation law that under
a macroscopic limit gives rise to diffusion. To be realistic
they should however share some features with the  Hamiltonian systems
that are more familiar and physically relevant. From this point of view
there is a lot missing from our analysis.

The first problem to understand is to go beyond the perturbative
analysis around $E=0$ (i.e. zero temperature). Then the equation
(\ref{lin}) picks also a driving term. 

The second unnatural assumption is the $E$-independence of the
$\theta$ dynamics. In a realistic model {rare configurations} of $E$ can
{slow down}  the  $\theta$ dynamics. Also the annealed system is probably not uniformly elliptic as we assumed and the random
drift can create traps in the environment with long lifetimes.

All these issues can and should be be studied with the renormalization group
approach sketched above.


\end{document}